\documentclass[prb,12pt]{revtex4}
\usepackage[english]{babel}
\usepackage[intlimits]{amsmath}
\usepackage[dvips]{graphicx}
\usepackage{epsfig}

\begin{document}

\begin{titlepage}

\title{Continuous-Time Random Walks at All Times}
\author{Anatoly B. Kolomeisky}
\affiliation{Department of Chemistry, Rice University, Houston, TX 77005-1892}

\begin{abstract}
Continuous-time random walks (CTRW) play important role in understanding of a wide range of phenomena. However, most theoretical studies of these models concentrate only on stationary-state dynamics. We present a new theoretical approach, based on generalized master equations picture, that allowed us to obtain explicit expressions for Laplace transforms for all dynamic quantities for different CTRW models. This theoretical method leads to the effective description of CTRW at all times. Specific calculations are performed for homogeneous, periodic models  and for CTRW with irreversible detachments. The approach to stationary states for CTRW is analyzed. Our results are also used to analyze generalized fluctuations theorem.

\end{abstract}

\maketitle

\end{titlepage}

\section{Introduction}

Continuous-time random walks (CTRW) are discreet models where particle transitions between different states  are controlled by random waiting-time distribution functions.\cite{montroll65,montroll73,landman77} CTRW is a powerful tool in studying dynamic processes in chemistry, physics, biology, social sciences and economics.\cite{weiss_book,hughes_book,rudnick_book,metzler00,zaslavsky02,econoCTRW} These models are especially convenient for investigation transport processes where deviations from the classical diffusion picture are observed.\cite{montroll65,montroll73,metzler00} Recently, CTRW have been successfully utilized for analyzing single-molecule enzyme kinetics,\cite{min05} dynamics of motor proteins,\cite{kolomeisky00,AR} and transport of metabolites across biological channels.\cite{dagdug09}

The majority of theoretical studies of CTRW address only dynamics at large times, i.e., the stationary-state properties.\cite{montroll65,weiss_book} But one might question the applicability of these results for understanding different processes since it is frequently not clear if the studied system reached the steady-state conditions. Strong advances in experimental techniques, especially in single-molecule biophysics and cellular transport, have provided us with  a comprehensive full-time description of many complex phenomena. It is reasonable to suggest that a theoretical method that permits a calculation of dynamic properties of the system at all times will provide a much better understanding of underlying mechanisms. The original work on CTRW\cite{montroll65} suggested a combined Laplace-Fourier transforms as a way to describe dynamics at all times. But these results, although formally correct, are technically impossible to apply for analyzing real systems. Recently, Berezhkovskii and Weiss\cite{BW08} introduced a new elegant method of computing exactly Laplace transforms of dynamics properties of CTRW. They successfully analyzed first two moments of the displacement of the random walker, the asymptotic behavior of the moments at large times, and the effective diffusion constant. However, this method has been developed only for homogeneous CTRW with the same set of waiting-time distribution functions at each site, and it is not clear how to extend it for more complex CTRW models.

In this paper we develop an alternative approach for calculating dynamics properties in complex CTRW models at all times. Utilizing the generalized master-equations description\cite{kenkre73,kolomeisky00} it is shown how  all Laplace transforms can be obtained explicitly for several complex CTRW model.

\section{Homogeneous CTRW}

To proceed, consider first the sequential CTRW kinetic model as shown in F1g. 1. The dynamics of the random walker is specified by waiting-time distribution functions $\psi_{j}^{\pm}(t)$  and $\psi_{j}^{\delta}(t)$. We define $\psi_{j}^{+}(t)dt$ as the probability of jumping one step forward from the state $j$ between times $t$ and $t+dt$ after arriving to the state $j$. Similarly,  $\psi_{j}^{-}(t)dt$ is the corresponding probability to move backward, while $\psi_{j}^{\delta}(t) dt$ determines the probability of irreversible detachment (or death) from the state $j$ at the same time interval. To illustrate our theoretical method, we will first analyze in detail the simplest homogeneous CTRW model with $\psi_{j}^{\pm}(t)=\psi^{\pm}(t)$ to be independent of the state $j$ and without detachments [$\psi_{j}^{\delta}(t)=0$ for all $j$].

Our theory is based on the crucial observation, presented first by Landman, Montroll and Shlesinger in 1977,\cite{landman77} that the probability $P_{n}(t)$ of finding the random walker at the site $n$ at time $t$ (assuming that at $t=0$ it started at $n=0$) is governed by a {\it generalized master equation}
\begin{equation}\label{master_eq}
\frac{d P_{n}(t)}{dt}=\int_{0}^{t} \left\{ \varphi^{+}(\tau) P_{n-1}(t-\tau) + \varphi^{-}(\tau) P_{n+1}(t-\tau)  - [\varphi^{+}(\tau) + \varphi^{-}(\tau)] P_{n}(t-\tau)  \right\} d\tau,
\end{equation}
where waiting-time rate distributions $\varphi^{\pm}(t)$\cite{kolomeisky00} are related to waiting-time distribution functions via Laplace transforms,
\begin{equation}\label{transform}
\widetilde{\varphi}^{\pm}(s)=\frac{s \widetilde{\psi}^{\pm}(s)}{1-\widetilde{\psi}(s)},
\end{equation}
with $\widetilde{\psi}(s)=\widetilde{\psi}^{+}(s)+\widetilde{\psi}^{-}(s)$. Performing Laplace transformation to the generalized master equation [Eq.(\ref{master_eq})] leads to
\begin{equation}
\left[s+\widetilde{\varphi}^{+}(s)+\widetilde{\varphi}^{-}(s)\right] \widetilde{P}_{n}(s)=\delta_{n,0}+ \widetilde{\varphi}^{+}(s)\widetilde{P}_{n-1}(s) + \widetilde{\varphi}^{-}(s)\widetilde{P}_{n+1}(s). 
\end{equation}
After introducing new variables, $a \equiv s+\widetilde{\varphi}^{+}(s)+\widetilde{\varphi}^{-}(s)$, $b \equiv \widetilde{\varphi}^{+}(s)$ and $c \equiv \widetilde{\varphi}^{-}(s)$, this equation can be written as
\begin{equation}\label{eqP}
a \widetilde{P}_{n}(s)=\delta_{n,0}+ b \widetilde{P}_{n-1}(s) + c \widetilde{P}_{n+1}(s). 
\end{equation}
We can define a new function, $\widetilde{R}_{n}(s)$, such that
\begin{equation}
\widetilde{P}_{n}(s) = \left(\frac{b}{c}\right)^{n/2} \widetilde{R}_{n}(s),
\end{equation}
and after substituting this relation into Eq. (\ref{eqP}) we obtain
\begin{equation}\label{eqR}
a \widetilde{R}_{n}(s)=\left(\frac{c}{b}\right)^{n/2} \delta_{n,0}+ \sqrt{bc}[\widetilde{R}_{n-1}(s) + \widetilde{R}_{n+1}(s)]. \end{equation}
This equation can be solved by looking for a solution in the form, $\widetilde{R}_{n}(s)=\widetilde{R}_{0}(s) x^{|n|}$, and after substituting into Eq. (\ref{eqR}) it yields
\begin{equation}
x=\frac{a \pm \sqrt{a^{2}-4 bc}}{2 \sqrt{bc}},
\end{equation}
and
\begin{equation}
\widetilde{R}_{0}(s)=1/\sqrt{a^{2}-4 bc}.
\end{equation}
Combining Eqs. (5)-(7), one can show that the final expression for the Laplace transform of the probability function is given by
\begin{equation}
\widetilde{P}_{n}(s) = \left(\frac{b}{c}\right)^{n/2} \left( \frac{ a- \sqrt{a^{2}-4 bc}}{2 \sqrt{bc}}\right)^{|n|} \frac{1}{\sqrt{a^{2}-4 bc}},
\end{equation}
or
\begin{eqnarray}\label{eq_Pn_phi}
& &\widetilde{P}_{n}(s) = \left(\frac{\widetilde{\varphi}^{+}(s)}{\widetilde{\varphi}^{-}(s)} \right)^{n/2} \left( \frac{ s+\widetilde{\varphi}^{+}(s)+\widetilde{\varphi}^{-}(s) - \sqrt{[s+\widetilde{\varphi}^{+}(s)+\widetilde{\varphi}^{-}(s)]^{2}-4 \widetilde{\varphi}^{+}(s)\widetilde{\varphi}^{-}(s)}}{2 \sqrt{\widetilde{\varphi}^{+}(s)\widetilde{\varphi}^{-}(s)}} \right)^{|n|} \times \nonumber \\
& & \frac{1}{\sqrt{[s+\widetilde{\varphi}^{+}(s)+\widetilde{\varphi}^{-}(s)]^{2}-4 \widetilde{\varphi}^{+}(s) \widetilde{\varphi}^{-}(s)}}.
\end{eqnarray}
It can also be written in terms of waiting-time distribution functions by utilizing Eq. (\ref{transform}),
\begin{eqnarray}\label{eq_Pn}
& &\widetilde{P}_{n}(s) = \left(\frac{\widetilde{\psi}^{+}(s)}{\widetilde{\psi}^{-}(s)} \right)^{n/2} \left( \frac{ 2\sqrt{\widetilde{\psi}^{+}(s)\widetilde{\psi}^{-}(s)}}{1+ \sqrt{1-4\widetilde{\psi}^{+}(s)\widetilde{\psi}^{-}(s)}} \right)^{|n|} \times \nonumber \\
& & \frac{1-\widetilde{\psi}(s)}{s\sqrt{1-4\widetilde{\psi}^{+}(s)\widetilde{\psi}^{-}(s)}}.
\end{eqnarray}
This is exactly the expression for the Laplace transform of the probability distribution function obtained in Ref.\cite{BW08} by a different approach.

These equations provide a direct way of analyzing dynamics of CTRW at all times. It can be shown by calculating explicitly first two moments of the motion. Defining $\langle n(t) \rangle$ as the average position of the random walker at time $t$, the following expression for the corresponding Laplace transform can be found
\begin{equation}\label{eqns}
\langle \widetilde{n}(s) \rangle= \sum_{n=-\infty}^{\infty} n \widetilde{P}_{n}(s) = \frac{b-c}{(a-b-c)^{2}}= \frac{\widetilde{\varphi}^{+}(s)-\widetilde{\varphi}^{-}(s)}{s^{2}}.
\end{equation}
Similar calculations for the Laplace transform of the second moment produce
\begin{eqnarray}\label{eqn2s}
& &\langle \widetilde{n}^{2}(s) \rangle= \sum_{n=-\infty}^{\infty} n^{2} \widetilde{P}_{n}(s) = \frac{(b+c)(a+b+c)-8bc}{(a-b-c)^{3}}= \nonumber \\
& & \frac{ \left[ \widetilde{\varphi}^{+}(s)+\widetilde{\varphi}^{-}(s) \right] \left[s+2\widetilde{\varphi}^{+}(s)+2\widetilde{\varphi}^{-}(s)\right] - 8 \widetilde{\varphi}^{+}(s)\widetilde{\varphi}^{-}(s)}{s^{3}}.
\end{eqnarray}

At large times the dynamic behavior of the first and second moments can be found by considering the limit of $s \rightarrow 0$. Expanding Laplace transforms of waiting-time rate distributions it can be shown that\cite{kolomeisky00}
\begin{eqnarray}
&\widetilde{\varphi}^{+}(s)& \simeq u + g^{+}s + \cdots, \nonumber \\
&\widetilde{\varphi}^{-}(s)& \simeq w + g^{-}s + \cdots,
\end{eqnarray}  
where $u=\widetilde{\varphi}^{+}(s=0)$ and $w=\widetilde{\varphi}^{-}(s=0)$ are effective transition rates;\cite{kolomeisky00} and $g^{\pm}=\frac{d\widetilde{\varphi}^{\pm}}{ds}|_{s=0}$. Substituting these expansions into Eqs. (\ref{eqns}) and (\ref{eqn2s}), one can find that
\begin{equation}
 \langle n(t) \rangle \simeq (u-w) t +(g^{+}-g^{-}),
\end{equation}
and 
\begin{equation}
 \langle n^{2}(t) \rangle \simeq (u-w)^{2} t^{2} +[(u+w)+4(u-w)(g^{+}-g^{-})]t+2(g^{+}-g^{-})^{2}+(g^{+}+g^{-}).
\end{equation}  
It is interesting to compare the characteristic times after which the dynamic property achieves its stationary-state behavior. For the first and for the second  moments these characteristic times are
\begin{equation}
t_{1}^{*} \simeq \frac{g^{+}-g^{-}}{u-w},
\end{equation}
and 
\begin{equation}
t_{2}^{*} \simeq \frac{u+w+4(u-w)(g^{+}-g^{-})}{(u-w)^{2}},
\end{equation}
respectively.
Then, because $t_{2}^{*} > t_{1}^{*}$, for some time interval the first moment can already be in the stationary state, while the second moment have  not reach it yet. The application of stationary-state formalism for analyzing CTRW  in this regime might lead to wrong description of its dynamics.  

From large-time behavior of the moments we could also compute other important dynamic properties such as the effective drift velocity $V$ and the effective diffusion constant $D$. The velocity is given by
\begin{equation}
V=\lim_{t \rightarrow \infty} \frac{ d \langle n(t) \rangle }{dt}=u-w,
\end{equation}
The corresponding expression for the diffusion constant is
\begin{equation}
D=\frac{1}{2} \lim_{t \rightarrow \infty} \frac{ d \langle n^{2}(t) \rangle - \langle n(t) \rangle^{2}}{dt}=\frac{1}{2}(u+w)+(u-w)(g^{+}-g^{-}).
\end{equation}
These equations reproduce stationary-state results obtained earlier for periodic $N$-state sequential CTRW with $N=1$ (the homogeneous case).\cite{kolomeisky00}

\section{CTRW with Irreversible Detachments}

The advantage of using this theoretical approach for computing dynamic properties at all times is the fact that it can be easily adopted for more complex CTRW models. Let us show this by considering CTRW model with irreversible detachments that presented in Fig. 1. We will again assume that  $\psi_{j}^{\pm}(t)=\psi^{\pm}(t)$ while the detachment waiting-time distribution functions are $\psi_{j}^{\delta}(t)=\psi^{\delta}(t) \ne 0$. This system is also homogeneous.

The probability to find the random walker at the site $n$ at time $t$ if at $t=0$ it was at the origin is again governed by the corresponding generalized master equation,\cite{landman77,kolomeisky00}
\begin{equation}\label{master_eqdetach}
\frac{d P_{n}(t)}{dt}=\int_{0}^{t}  \varphi^{+}(\tau) P_{n-1}(t-\tau) + \varphi^{-}(\tau) P_{n+1}(t-\tau)  - [\varphi^{+}(\tau) + \varphi^{-}(\tau)+\varphi^{\delta}(\tau)] P_{n}(t-\tau)   d\tau,
\end{equation}
where $\varphi^{+}(t)$, $\varphi^{-}(t)$ and $\varphi^{\delta}(t)$ are waiting-time rate distributions for moving forward, backward and to detach, respectively. They can be expressed in terms of original waiting-time distribution functions,
\begin{equation}\label{transform_detach}
\widetilde{\varphi}^{i}(s)=\frac{s \widetilde{\psi}^{i}(s)}{1-\widetilde{\psi}(s)},
\end{equation}
with $\widetilde{\psi}(s)=\widetilde{\psi}^{+}(s)+\widetilde{\psi}^{-}(s)+\widetilde{\psi}^{\delta}(s)$ and $i=+$, $-$ or $\delta$. In this system the probability is not conserved, and it is convenient to define a new function $Q_{n}(t)$ defined as\cite{kolomeisky00}
\begin{equation}\label{PnQn}
P_{n}(t)=e^{-\lambda t} Q_{n}(t).
\end{equation}
The function $Q_{n}(t)$ has a meaning of the survival probability of reaching the site $n$ at time $t$, and the parameter $\lambda$ is an effective detachment rate for the random walker. Then the generalized master equation should be modified as
\begin{eqnarray}\label{master_eqQ}
\frac{d Q_{n}(t)}{dt} & = & \int_{0}^{t} \left\{ \varphi^{+}(\tau) e^{\lambda \tau} Q_{n-1}(t-\tau) + \varphi^{-}(\tau) e^{\lambda \tau} Q_{n+1}(t-\tau)  \right.-  \nonumber\\
& & \left. [\varphi^{+}(\tau) + \varphi^{-}(\tau)+\varphi^{\delta}(\tau)] e^{\lambda \tau} Q_{n}(t-\tau) \right\} d\tau +\lambda Q_{n}(t).
\end{eqnarray}
Then after Laplace transformation this equation produces,
\begin{eqnarray}\label{eq_Q}
& & \left[s+\widetilde{\varphi}^{+}(s-\lambda)+\widetilde{\varphi}^{-}(s-\lambda)+\widetilde{\varphi}^{\delta}(s-\lambda) -\lambda \right] \widetilde{Q}_{n}(s)= \nonumber \\
& & \delta_{n,0}+ \widetilde{\varphi}^{+}(s-\lambda)\widetilde{Q}_{n-1}(s) + \widetilde{\varphi}^{-}(s-\lambda)\widetilde{Q}_{n+1}(s). 
\end{eqnarray}
To simplify notations we again define,
\begin{eqnarray}
a & \equiv & s+\widetilde{\varphi}^{+}(s-\lambda)+\widetilde{\varphi}^{-}(s-\lambda)+\widetilde{\varphi}^{\delta}(s-\lambda) -\lambda , \nonumber \\
b & \equiv & \widetilde{\varphi}^{+}(s-\lambda), \nonumber \\
c & \equiv & \widetilde{\varphi}^{-}(s-\lambda).
\end{eqnarray}
As shown in the Sec. II,  Eq. (\ref{eq_Q}) can be solved to yield the expression for the Laplace transform for the survival probability,
\begin{equation}\label{eqQns}
\widetilde{Q}_{n}(s) = \left(\frac{b}{c}\right)^{n/2} \left( \frac{ a- \sqrt{a^{2}-4 bc}}{2 \sqrt{bc}}\right)^{|n|} \frac{1}{\sqrt{a^{2}-4 bc}}.
\end{equation}
The corresponding expression  for the probability function $P_{n}(t)$ can be easily obtained from Eq. (\ref{PnQn}),
\begin{equation}
\widetilde{P}_{n}(s) = \widetilde{Q}_{n}(s+\lambda). 
\end{equation}

Explicit equations for Laplace transforms can be used to calculate behavior of all relevant dynamic properties at all times. In this system the first moment is defined as
\begin{equation}
 \langle n(t) \rangle =\frac{\sum_{n} n P_{n}(t)}{\sum_{n} P_{n}(t)}= \frac{\sum_{n} n Q_{n}(t)}{\sum_{n} Q_{n}(t)}.
\end{equation}
It can be shown that the Laplace transform of the first moment is equal to
\begin{equation}
\langle \widetilde{n}(s) \rangle= \frac{\sum_{n} n \widetilde{Q}_{n}(s)}{s \sum_{n}  \widetilde{Q}_{n}(s)}.
\end{equation}
Utilizing Eq. (\ref{eqQns}) it can be calculated that
\begin{equation}
\langle \widetilde{n}(s) \rangle= \frac{b-c}{s(a-b-c)}= \frac{\widetilde{\varphi}^{+}(s-\lambda)-\widetilde{\varphi}^{-}(s-\lambda)}{s[s+\widetilde{\varphi}^{\delta}(s-\lambda)-\lambda]}.
\end{equation}
We can also perform similar analysis for the second moment, and it  yields
\begin{eqnarray}
& &\langle \widetilde{n}^{2}(s) \rangle = \frac{(b+c)(a+b+c)-8bc}{s(a-b-c)^{2}}= \left[ \widetilde{\varphi}^{+}(s-\lambda)+\widetilde{\varphi}^{-}(s-\lambda) \right] \times \nonumber \\
& & \frac{\left[ s+2\widetilde{\varphi}^{+}(s-\lambda)+2\widetilde{\varphi}^{-}(s-\lambda) + \widetilde{\varphi}^{\delta}(s-\lambda) -\lambda \right] - 8 \widetilde{\varphi}^{+}(s-\lambda)\widetilde{\varphi}^{-}(s-\lambda)}{s \left[s+\widetilde{\varphi}^{-}(s-\lambda)-\lambda \right]^{2}}.
\end{eqnarray}

The stationary-state behavior of the first and second moments can be found by expanding at small $s$ expressions for $\widetilde{\varphi}^{+}(s-\lambda)$, $\widetilde{\varphi}^{-}(s-\lambda)$ and $\widetilde{\varphi}^{\delta}(s-\lambda)$:
\begin{eqnarray}
&\widetilde{\varphi}^{+}(s-\lambda)      & \simeq u + g^{+}s + \frac{1}{2} h^{+} s^{2}  \cdots, \nonumber \\
&\widetilde{\varphi}^{-}(s-\lambda)      & \simeq w + g^{-}s + \frac{1}{2} h^{-} s^{2}  \cdots, \nonumber \\
&\widetilde{\varphi}^{\delta}(s-\lambda) & \simeq \lambda + g^{\delta}s + \frac{1}{2} h^{\delta} s^{2} + \cdots,
\end{eqnarray} 
Using these expansions leads us to the following expressions for stationary-state behavior of the first and second moments,
\begin{equation}
 \langle n(t) \rangle \simeq \frac{(u-w)}{(1+g^{\delta})} t +\frac{(g^{+}-g^{-})}{(1+g^{\delta})}-\frac{(u-w)h^{\delta}}{2(1+g^{\delta})^{2}},
\end{equation}
\begin{equation}
 \langle n^{2}(t) \rangle \simeq \frac{(u-w)^{2}}{(1+g^{\delta})^{2}}t^{2} + \left[\frac{(u+w)}{(1+g^{\delta})}+\frac{4(u-w)(g^{+}-g^{-})}{(1+g^{\delta})^{2}} - \frac{2(u-w)^{2} h^{\delta}}{(1+g^{\delta})^{3}}\right] t.
\end{equation}  
From these results we can easily compute the effective drift velocity,
\begin{equation}
V= \frac{(u-w)}{(1+g^{\delta})};
\end{equation}
and the effective diffusion constant,
\begin{equation}
D=\frac{(u+w)}{2(1+g^{\delta})}+ \frac{(u-w)(g^{+}-g^{-})}{(1+g^{\delta})^{2}}-\frac{(u-w)^{2} h^{\delta}}{2(1+g^{\delta})^{3}}.
\end{equation}
These equations are identical to the steady-state expressions for $V$ and $D$ obtained earlier via a different method.\cite{kolomeisky00,footnote}

\section{Periodic CTRW}

This theoretical method can also be extended for periodic CTRW. Let us consider the simplest non-trivial periodic system as shown in Fig. 1 with $\psi_{j}^{\pm}(t)=\psi_{0}^{\pm}(t)$ for $j$ even and $\psi_{j}^{\pm}(t)=\psi_{1}^{\pm}(t)$ for $j$ add. This corresponds to the periodic $N$-state CTRW model with $N=2$. Irreversible detachments  will be neglected in this case, i.e., $\psi_{j}^{\delta}(t)=0$ for all $j$.

The temporal evolution of the system is described by a system of generalized master equations,
\begin{eqnarray}\label{master_eq_periodic}
& \frac{d P_{2n}(t)}{dt}  & = \int_{0}^{t}  \varphi_{1}^{+}(\tau) P_{2n-1}(t-\tau) + \varphi_{1}^{-}(\tau) P_{2n+1}(t-\tau)  - [\varphi_{0}^{+}(\tau) + \varphi_{0}^{-}(\tau)] P_{2n}(t-\tau)   d\tau; \nonumber \\
& \frac{d P_{2n+1}(t)}{dt} & = \int_{0}^{t}  \varphi_{0}^{+}(\tau) P_{2n}(t-\tau) + \varphi_{0}^{-}(\tau) P_{2n+2}(t-\tau)  - [\varphi_{1}^{+}(\tau) + \varphi_{1}^{-}(\tau)] P_{2n+1}(t-\tau)   d\tau.
\end{eqnarray}
As before, waiting-time rate distributions $\varphi_{i}^{\pm}(t)$  are closely connected to waiting-time distribution functions $\psi_{i}^{\pm}(t)$ ($i=0$, $1$)  as given by
\begin{equation}
\widetilde{\varphi}_{i}^{\pm}(s)=\frac{s \widetilde{\psi}_{i}^{\pm}(s)}{1-\widetilde{\psi}_{i}(s)},
\end{equation}
for $i=0$ or $1$, and with $\widetilde{\psi}_{i}(s)=\widetilde{\psi}_{i}^{+}(s)+\widetilde{\psi}_{i}^{-}(s)$. Laplace transformations change generalized master equations into
\begin{eqnarray}
\left[s+\widetilde{\varphi}_{0}^{+}(s)+\widetilde{\varphi}_{0}^{-}(s)\right] \widetilde{P}_{2n}(s) & = & \delta_{n,0}+ \widetilde{\varphi}_{1}^{+}(s)\widetilde{P}_{2n-1}(s) + \widetilde{\varphi}_{1}^{-}(s)\widetilde{P}_{2n+1}(s); \nonumber \\
\left[s+\widetilde{\varphi}_{1}^{+}(s)+\widetilde{\varphi}_{1}^{-}(s)\right] \widetilde{P}_{2n+1}(s) & = & \delta_{n,0}+ \widetilde{\varphi}_{0}^{+}(s)\widetilde{P}_{2n}(s) + \widetilde{\varphi}_{0}^{-}(s)\widetilde{P}_{2n+2}(s).
\end{eqnarray}
After introducing the  auxiliary functions
\begin{equation}
a_{i}  \equiv  s+\widetilde{\varphi}_{i}^{+}(s)+\widetilde{\varphi}_{i}^{-}(s), \quad b_{i}  \equiv  \widetilde{\varphi}_{i}^{+}(s) \quad  c_{i}  \equiv  \widetilde{\varphi}_{i}^{-}(s),
\end{equation}
for $i=0$ or $1$; and defining $P_{2n}(t)=Q_{n}(t)$ and $P_{2n+1}(t)=R_{n}(t)$, we obtain
\begin{eqnarray}
a_{0} \widetilde{Q}_{n}(s)=\delta_{n,0}+ b_{1} \widetilde{R}_{n-1}(s) + c_{1} \widetilde{R}_{n}(s), \nonumber \\
a_{1} \widetilde{R}_{n}(s)= b_{0} \widetilde{Q}_{n}(s) + c_{0} \widetilde{Q}_{n+1}(s).
\end{eqnarray}
Combining these two equations we can write
\begin{equation}\label{eqN2}
A \widetilde{Q}_{n}(s)= a_{1}\delta_{n,0}+ B \widetilde{Q}_{n-1}(s) + C \widetilde{Q}_{n+1}(s),
\end{equation}
where
\begin{equation}
A=a_{0} a_{1}-b_{1}c_{0}-b_{0}c_{1}, \quad B=b_{0}b_{1}, \quad C=c_{0}c_{1}.
\end{equation}
Eq. (\ref{eqN2}) can be solved to produce the Laplace transform for the probability on even sites,
\begin{eqnarray}
& &\widetilde{Q}_{n}(s)  =  \left(\frac{B}{C}\right)^{n/2} \left( \frac{ A- \sqrt{A^{2}-4 BC}}{2 \sqrt{BC}}\right)^{|n|} \frac{a_{1}}{\sqrt{A^{2}-4 BC}} = \left(\frac{b_{0}b_{1}}{c_{0}c_{1}}\right)^{n/2} \times \nonumber \\
& &  \left( \frac{a_{0} a_{1}-b_{1}c_{0}-b_{0}c_{1}  - \sqrt{(a_{0} a_{1}-b_{1}c_{0}-b_{0}c_{1})^{2}-4 b_{0}b_{1}c_{0}c_{1}}}{2 \sqrt{b_{0}b_{1}c_{0}c_{1}}}\right)^{|n|}  \times \nonumber \\
& & \frac{a_{1}}{\sqrt{(a_{0} a_{1}-b_{1}c_{0}-b_{0}c_{1})^{2}-4 b_{0}b_{1}c_{0}c_{1}}}.
\end{eqnarray}
Simultaneously, it can be shown that for the odd sites,
\begin{equation}
\widetilde{R}_{n}(s)=\frac{b_{0}}{a_{1}} \widetilde{Q}_{n}(s) + \frac{c_{0}}{a_{1}} \widetilde{Q}_{n+1}(s).
\end{equation}

Now we can calculate the dynamic behavior of the first and second moments. For the first moment, 
\begin{equation}
 \langle n(t) \rangle =\sum_{n} n P_{n}(t) = \sum_{n} 2n Q_{n}(t) + \sum_{n} (2n+1)R_{n}(t).
\end{equation}
Then using explicit expressions for $\widetilde{Q}_{n}(s)$ and $\widetilde{R}_{n}(s)$, our calculations yield
\begin{equation}
\langle \widetilde{n}(s) \rangle = \frac{b_{0}-c_{0}}{s(s+b_{0}+b_{1}+c_{0}+c_{1})}+ \frac{2(b_{0}b_{1}-c_{0}c_{1})}{s^{2}(s+b_{0}+b_{1}+c_{0}+c_{1})}.
\end{equation}
Similar calculations for the Laplace transform of the second moment are quite tedious, but the corresponding expression is relatively compact,
\begin{eqnarray}
& &\langle \widetilde{n}^{2}(s) \rangle  = \frac{4(b_{0}b_{1}+c_{0}c_{1})}{s^{2}(s+b_{0}+b_{1}+c_{0}+c_{1})} + \frac{8(b_{0}b_{1}-c_{0}c_{1})^{2}}{s^{3}(s+b_{0}+b_{1}+c_{0}+c_{1})^{2}} +\nonumber \\
& & \frac{4(b_{0}b_{1}-c_{0}c_{1})(b_{0}-c_{0})}{s^{2}(s+b_{0}+b_{1}+c_{0}+c_{1})^{2}} + \frac{(b_{0}+c_{0})}{s(s+b_{0}+b_{1}+c_{0}+c_{1})}.
\end{eqnarray}
At large times, the dynamics can be understood if we use expansions of waiting-time rate distribution functions at $s \rightarrow 0$,\cite{kolomeisky00}
\begin{eqnarray}
&b_{0}=\widetilde{\varphi}_{0}^{+}(s)& \simeq u_{0} + g_{0}^{+}s + \cdots, \nonumber \\
&b_{1}=\widetilde{\varphi}_{1}^{+}(s)& \simeq u_{1} + g_{1}^{+}s + \cdots, \nonumber \\
&c_{0}=\widetilde{\varphi}_{0}^{-}(s)& \simeq w_{0} + g_{0}^{-}s + \cdots, \nonumber \\
&c_{1}=\widetilde{\varphi}_{1}^{-}(s)& \simeq w_{1} + g_{1}^{-}s + \cdots.
\end{eqnarray} 
Then at $t \rightarrow \infty$ we have for the first moment,
\begin{eqnarray}
\langle n(t) \rangle & \simeq &\frac{2(u_{0}u_{1}-w_{0}w_{1})}{(u_{0}+u_{1}+w_{0}+w_{1})} t + \frac{2(u_{1}g_{0}^{+} + u_{0}g_{1}^{+})-2(w_{0}g_{1}^{-} + w_{1}g_{0}^{-})+(u_{0}-w_{0})}{(u_{0}+u_{1}+w_{0}+w_{1})} - \nonumber \\
& & \frac{2(u_{0}u_{1}-w_{0}w_{1})(1+g_{0}^{+}+g_{1}^{+}+g_{0}^{-}+g_{1}^{-})}{(u_{0}+u_{1}+w_{0}+w_{1})^{2}},
\end{eqnarray}
while for the second moment,
\begin{eqnarray}
 \langle n^{2}(t) \rangle & \simeq & \frac{4(u_{0}u_{1}-w_{0}w_{1})^{2}}{(u_{0}+u_{1}+w_{0}+w_{1})^{2}} t^{2} +\left[\frac{4(u_{0}u_{1}+w_{0}w_{1})}{(u_{0}+u_{1}+w_{0}+w_{1})} + \frac{4(u_{0}u_{1}-w_{0}w_{1})(u_{0}-w_{0})}{(u_{0}+u_{1}+w_{0}+w_{1})^{2}}  \right. \nonumber \\
& & + \frac{16(u_{1}g_{0}^{+} + u_{0}g_{1}^{+} - w_{0}g_{1}^{-} - w_{1}g_{0}^{-})(u_{0}u_{1}-w_{0}w_{1})}{(u_{0}+u_{1}+w_{0}+w_{1})^{2}} \nonumber \\
& & \left. - \frac{16(u_{0}u_{1}-w_{0}w_{1})^{2}(1+g_{0}^{+}+g_{1}^{+}+g_{0}^{-}+g_{1}^{-})}{(u_{0}+u_{1}+w_{0}+w_{1})^{3}}\right] t.
\end{eqnarray}  
These expressions allow us to derive the  stationary drift velocity,
\begin{equation}
V=\frac{2(u_{0}u_{1}-w_{0}w_{1})}{(u_{0}+u_{1}+w_{0}+w_{1})},
\end{equation}
while for the effective diffusion constant we have, 
\begin{eqnarray}
D & = &\frac{2(u_{0}u_{1}+w_{0}w_{1})}{(u_{0}+u_{1}+w_{0}+w_{1})} + \frac{4(u_{0}u_{1}-w_{0}w_{1})(u_{1}g_{0}^{+} + u_{0}g_{1}^{+} - w_{0}g_{1}^{-} - w_{1}g_{0}^{-})}{(u_{0}+u_{1}+w_{0}+w_{1})^{2}} \nonumber \\
& & - \frac{4(u_{0}u_{1}-w_{0}w_{1})^{2}(1+g_{0}^{+}+g_{1}^{+}+g_{0}^{-}+g_{1}^{-})}{(u_{0}+u_{1}+w_{0}+w_{1})^{3}}.
\end{eqnarray} 
These equations reproduce, as expected, known results for dynamic properties of CTRW at stationary-state conditions.\cite{kolomeisky00} Calculations for periodic CTRW models with $N>2$ can be accomplished following this approach.

\section{Generalized Fluctuation Theorem}

Berezhkovskii and Weiss\cite{BW08} introduced a generalized fluctuation theorem by considering the ratio $\frac{\widetilde{P}_{n}(s)}{\widetilde{P}_{-n}(s)}$. It was argued that it reduces to the conventional form of the fluctuation theorem\cite{gallavotti95,gaspard04,seifert05} under some conditions. For homogeneous CTRW it was found\cite{BW08} [see also our Eqs. (\ref{eq_Pn_phi}) and (\ref{eq_Pn})] that
\begin{equation}\label{fluct}
\frac{\widetilde{P}_{n}(s)}{\widetilde{P}_{-n}(s)}=\left[ \frac{\widetilde{\psi}^{+}(s)}{\widetilde{\psi}^{-}(s)} \right]^{n}=\left[ \frac{\widetilde{\varphi}^{+}(s)}{\widetilde{\varphi}^{-}(s)} \right]^{n}.
\end{equation}
It leads to the original fluctuation theorem result when $\frac{\psi^{+}(t)}{\psi^{-}(t)}$ is time-independent. Our calculations allow to extend this result for periodic $N$-state CTRW with $N=2$. Using Eqs. (41) and (45) we obtain,
\begin{equation}
\frac{\widetilde{P}_{2n}(s)}{\widetilde{P}_{-2n}(s)}=\left[ \frac{\widetilde{\psi}_{0}^{+}(s)\widetilde{\psi}_{1}^{+}(s)}{\widetilde{\psi}_{0}^{-}(s)\widetilde{\psi}_{1}^{-}(s)} \right]^{n}=\left[ \frac{\widetilde{\varphi}_{0}^{+}(s)\widetilde{\varphi}_{1}^{+}(s)}{\widetilde{\varphi}_{0}^{-}(s)\widetilde{\varphi}_{1}^{-}(s)} \right]^{n}.
\end{equation}
It is also interesting to consider the generalized fluctuation theorem for CTRW with irreversible detachments. In this case we obtain from Eqs. (27) and (28) the expression identical to Eq. (\ref{fluct}). This result is rather surprising, and it suggests that irreversible detachments do not affect the ratio of probabilities  for the forward and backward steps of the random walker, and consequently do not change statistics for occurrence of different fluctuations.

\section{Summary and Conclusions}

A new theoretical approach of computing dynamic properties of CTRW at all times is developed. The important critical step of the method is the observation that CTRW dynamics can be described by generalized master equations. It allowed us to compute exactly Laplace transforms of probability functions and all other relevant dynamic properties. The presented approach is flexible enough to describe complex CTRW systems. Specifically, we analyzed homogeneous and periodic homogeneous CTRW, as well as the model with irreversible detachments.   

We calculated explicitly how different CTRW models approach their stationary-state dynamics. All derived expressions for dynamic properties at steady states agree with available large-time results for CTRW obtained by different methods. Based on these observations it is argued that times to reach the steady-state conditions  might differ significantly for different dynamic properties, and one must be careful in applying stationary-state results for understanding mechanisms of processes described by CTRW models. We also analyzed the generalized fluctuation theorem, and it is found that irreversible detachments do not affect much the fluctuation dynamics. It will be interesting to extend this theoretical approach to analyze at all times more complex CTRW systems, such as models with branched states\cite{AR} and parallel-chain models.\cite{kolomeisky01,das09}

\section*{Acknowledgments}

The author would like to acknowledge the support from the Welch Foundation (Grant No. C-1559), and the U.S. National Science Foundation (Grant No. ECCS-0708765).

\newpage

\noindent {\bf Figure Captions:} \\
\vspace{5mm}

\noindent Fig. 1. A general scheme for continuous-time random walk (CTRW) models. $\psi_{j}^{+}(t)$, $\psi_{j}^{-}(t)$ and $\psi_{j}^{\delta}(t)$ are waiting-time distribution functions to step forward, backward or to dissociate irreversibly, correspondingly.

\newpage

\begin{figure}[ht]
\begin{center}
\unitlength 1in
\begin{picture}(3.0,4.0)
  \resizebox{3.375in}{2in}{\includegraphics{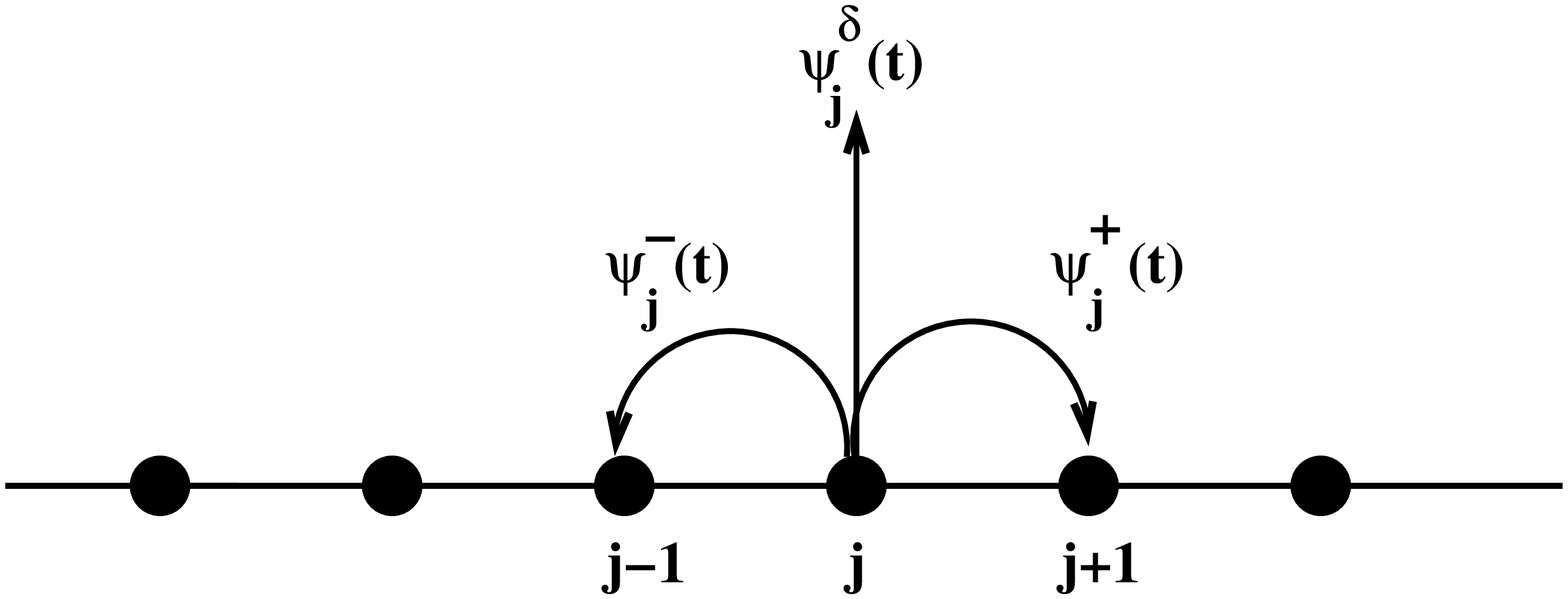}}
\end{picture}
\vskip 1in
 \begin{Large} Figure 1. Kolomeisky \end{Large}
\end{center}
\end{figure}


\begin{thebibliography}{99}
\bibitem{footnote} Note that in Eq. (45) of Ref.\cite{kolomeisky00} there must be a negative sign in front of the right side.
\end{thebibliography}
\end{document}